\documentclass[bibyear]{aa} 

\usepackage{graphicx}
\usepackage{txfonts}
\usepackage{lipsum}
\usepackage{subcaption}         
\usepackage{lscape}             
\usepackage{placeins}           
                                
\usepackage{txfonts}
\usepackage{graphicx}	
\usepackage{amsmath}	
\usepackage{mathrsfs}
\usepackage[breaklinks, colorlinks, citecolor=blue, urlcolor = blue]{hyperref}
\usepackage{pdflscape}
\usepackage{float}
\usepackage{here}
\usepackage[export]{adjustbox}
\usepackage{array}
\usepackage{natbib}
\bibpunct{(}{)}{;}{a}{}{,} 

\newcommand{\lx}{$L_{2-10\,\rm keV}$}
\newcommand{\lha}{$L_{\rm H\alpha}$}

\newcommand{\xmm}{XMM-\textit{Newton }}
\newcommand{\ha}{H$\alpha$}

\begin{document}

   \title{The X-ray weakness of little red dots and JWST-selected AGN: Comparison with local AGN in different accretion regimes}
   \titlerunning{LRDs, JWST-AGN and local AGN}
   \authorrunning{Tortosa et al.}



   \author{A. Tortosa\thanks{\email{alessia.tortosa@inaf.it}}\inst{1}
   \and C. Ricci \inst{2,3}
   \and P. Du \inst{4}
   \and G. Venturi \inst{5}
   \and L. C. Ho \inst{6,7}
   \and R. Li \inst{8}
   \and J.-M. Wang \inst{4,9,10}
   \and M. Berton \inst{11}
        }

   \institute{INAF - Osservatorio astronomico di Roma, Via Frascati 33, I-00040 Monte Porzio Catone, Italy.
   \and Department of Astronomy, University of Geneva, ch. d’Ecogia 16, 1290, Versoix, Switzerland.
   \and Instituto de Estudios Astrofìsicos, Facultad de Ingenierìa y Ciencias, Universidad Diego Portales, Av. Ejèrcito Libertador 441, Santiago, Chile.
   \and Key Laboratory for Particle Astrophysics, Institute of High Energy Physics, Chinese Academy of Sciences, 19B Yuquan Road, Beijing 100049, People’s Republic of China.
   \and Scuola Normale Superiore, Piazza dei Cavalieri 7, I-56126 Pisa, Italy.
   \and Kavli Institute for Astronomy and Astrophysics, Peking University, Beijing 100871, People’s Republic of China.
   \and Department of Astronomy, School of Physics, Peking University, Beijing 100871, People’s Republic of China.
   \and Max-Planck-Institut f\"ur extraterrestrische Physik, Giessenbachstra{\ss}e, 85748 Garching, Germany.
   \and School of Astronomy and Space Sciences, University of Chinese Academy of Sciences, 19A Yuquan road, Beijing 100049, People’s Republic of China.
   \and National Astronomical Observatory of China, 20A Datun Road, Beijing 100020, People’s Republic of China.
   \and European Southern Observatory (ESO), Alonso de Còrdova 3107, Casilla 19, Santiago, Chile.
     }

   \date{Received XX, 2025; accepted YY, 2025}

 
  \abstract
   {We investigate the origin of the observed X-ray weakness in high $z$ little red dots (LRDs) and other JWST-selected broad line active galactic nuclei (AGN) by comparing their X-ray and optical properties with those of a diverse sample of low $z$ AGN, including super-Eddington accreting massive black holes (SEAMBHs), narrow-line Seyfert 1 galaxies (NLS1s), and type~I AGN from large surveys (e.g. BASS, SDSS). Using a heterogeneous set of AGN samples spanning a wide range of redshift and accretion rates, we examine the relations between X-ray luminosity (\lx), broad \ha\ line luminosity (\lha), Eddington ratio ($\lambda_{\rm Edd}$), bolometric luminosity ($L_{\rm bol}$), and X-ray-to-bolometric luminosity correction ($\kappa_{\mathrm{bol,X}}$), and we explore whether high $z$ sources may represent analogues of local highly accreting systems. 
   
   While a few LRDs and JWST-selected AGN are consistent with the SEAMBH population in the \lx/\lha\ versus $\lambda_{\rm Edd}$ plane, most lie below it, suggesting either more extreme accretion conditions, suppressed coronal emission or heavy obscuration.
   
   We identify an anti-correlation between \lx/\lha\ and $\lambda_{\rm Edd}$ in the low $z$, high-$\lambda_{\rm Edd}$ subsample of sources, consistent with theoretical expectations of slim-disc accretion. We further show that, for SEAMBHs, \ha-based bolometric luminosities underestimate spectral energy distribution-based values even after dust correction, reinforcing the need for SED-based estimates.
   
   We find that SEAMBHs, LRDs, and JWST-selected AGN occupy a similar high-$\kappa_{\mathrm{bol,X}}$ regime, indicating that the relative deficit of X-ray emission compared to the bolometric output could potentially support the view that suppression of the hot corona emission is a common feature of highly accreting systems across cosmic time. However, the X-ray measurements of high $z$ sources are largely based on observed upper limits and generally do not account for heavy or Compton-thick obscuration, in which case the intrinsic \lx\ could be substantially higher than observed.
   
   Our results are consistent with the idea that the observed X-ray weakness of LRDs and JWST-selected AGN may be linked to the physics of highly accreting SMBHs, but alternative explanations, including heavy obscuration, systematics in BH mass estimates, or a combination of intrinsic coronal suppression and absorption, remain viable. Moreover, observational limitations at high $z$, including instrumental sensitivity and the steep X-ray spectra expected for highly accreting systems, likely further suppress the detected X-ray signal. Disentangling the roles of accretion physics and obscuration will require deeper, higher-resolution X-ray observations with next-generation facilities, which will be crucial for establishing whether these sources represent genuine high $z$ counterparts of local highly accreting AGN.}


   \maketitle

\section{Introduction}
\nolinenumbers
\label{sect:intro}
Recent observations with the \textit{James Webb Space Telescope} (JWST) have revealed a population of compact, high-redshift active galactic nuclei (AGN) \citep[e.g.]{Harikane_2023,Kocevski_2023,Yue_2024,Greene_2024,Maiolino2025}. Among these newly identified sources is a population of peculiar objects at $z\gtrsim4$, the so-called little red dots \citep[LRDs; see][for a recent review]{Inayoshi_Ho_2025}. The LRDs present intriguing characteristics that challenge, and could potentially improve, our understanding of early cosmic structures. They show a V-shaped spectral energy distribution (SED) with a very red rest-frame optical colour and a blue UV colour \citep[e.g.][]{Greene_2024,Harikane_2023,Kocevski_2023,Labbe_2025,Matthee_2024}. The origin of LRDs is still uncertain. A significant fraction of spectroscopically confirmed LRDs (approximately 60\%) display broad \ha\ line emission, which could be consistent with type~I AGN activity in LRD centres \citep{2025A&A...702A..57H,Greene_2024}. However, alternative explanations for the presence of such features, such as outflows or high central stellar densities have also been proposed \citep{Harikane_2023,Barro24,PerezGonzalez24,Labbe_2025}. The remaining fraction of LRDs with no broad \ha\ detection may host obscured or low-luminosity AGN, or be powered by alternative mechanisms.
Importantly, LRDs are not simply a subset of JWST-selected AGN; only about 10--30\% of JWST-discovered AGN satisfy the compactness and colour criteria used to define the LRD population \citep[e.g.][]{2025ApJ...979..138H}. At the same time, spectroscopic studies have indicated that broad Balmer emission is detected in the majority of LRDs where data quality and spectral coverage are sufficient, suggesting that the intrinsic fraction of LRDs hosting broad-line AGN may be substantially higher than implied by current detection rates \citep{2025A&A...702A..57H}. In this work, we focus on the LRDs with spectroscopic AGN signatures, for which broad \ha\ measurements are available.

One of the most striking observational features of LRDs and JWST-selected AGN is their undetected X-ray emission, with upper limits often lying well below the expectations from standard AGN scaling relations, such as the one between the X-ray luminosity (\lx) and the luminosity of the broad component of the H$\alpha$ emission line (\lha) \citep{Yue_2024}. The relative balance between optical/UV and X-ray emission in AGN is understood to reflect the energetic coupling between the accretion disc and the hot corona. Thermal photons produced in the disc are Compton up-scattered by energetic electrons in the corona, generating the hard X-ray continuum, while the ionising radiation field from the inner accretion flow photoionises the broad-line region (BLR), producing recombination lines such as \ha\ \citep[e.g.][]{1980A&A....86..121S,1993ApJ...413..507H}. As a result, empirical correlations between optical/UV and X-ray tracers of accretion power are expected in radiatively efficient systems, and departures from these relations may signal changes in the structure or radiative efficiency of the inner accretion flow \citep[e.g.][]{Lusso2016,Netzer2019}.

The detected X-ray weakness raises important questions about the physical conditions in high $z$ AGN mainly whether these objects are heavily obscured or intrinsically lack a luminous corona. Interestingly, similar behaviour has been observed in local super-Eddington accreting massive black holes \citep[SEAMBHs;][]{Wang2014,Du2014,Du2015,Du2018,Tortosa22,Tortosa2023}, which also show suppressed X-ray emission relative to optical broad-line luminosities. Theoretical models predict that the structure and radiative properties of the disc-corona system should depend strongly on the accretion rate. At high Eddington ratios, the inner accretion flow is expected to become geometrically thick, entering the so-called slim-disc regime, where photon trapping, advection, and radiation pressure significantly modify the energy balance of the flow \citep{1988ApJ...332..646A,Sadowski2016}. The enhanced radiation density in these systems can increase Compton cooling of the corona and drive powerful disc winds, potentially reducing the efficiency of hard X-ray production or altering the geometry of the X-ray emitting region \citep{Jiang2019,2020ApJ...894..141I}. Several of the properties exhibited by LRDs and JWST-selected AGN, including their compactness and high inferred accretion rates, bear a striking resemblance to those of local narrow-line Seyfert 1 galaxies (NLS1s;, \citealp{1994ApJ...435L.125M,2002A&A...388..771C,Collin_2004}; for a recent review see \citealt{Berton2025} and references therein). 
Among NLS1s, the sources belonging to the SEAMBHs sample have been proposed as local analogues of early black hole growth episodes, making them a valuable comparison point for understanding LRDs. Both populations appear to show weak X-ray emission and exhibit low \lx/\lha\ values, though they occupy vastly different epochs in cosmic history. However, recent works argue that these systems are embedded within a compact, dense gas cocoon \citep{Naidu25,Degraaff25,Inayoshi25,Inayoshi25a}, a configuration that can reproduce their Balmer break signatures and absorption features. This configuration may also be responsible for their suppressed X-ray emission.

In this work, we aim to place LRDs and JWST-selected AGN in a broader AGN context by comparing their X-ray and broad \ha\ line emission properties with those of a diverse sample of low-redshift AGN, including: SEAMBHs, NLS1s, and AGN from large spectroscopic surveys such as the Swift/BAT AGN Spectroscopic Survey (BASS\footnote{\url{www.bass-survey.com}}) and the Sloan Digital Sky Survey \citep[SDSS;][]{York_2000}. 

By examining the relationship between the X-ray-to-\ha\ luminosity ratio and the Eddington ratio across redshift, we assess whether X-ray weak high $z$ AGN are consistent with being high $z$ analogues of SEAMBHs or if they have a different nature. Our goal is to better constrain the accretion properties of these sources and to test whether the empirical relations established for low $z$ AGN remain valid in the early Universe. Standard cosmological parameters (H=70\,km\,s$^{-1} \rm Mpc^{-1}$, $\Omega_{\Lambda}$=0.73 and $\Omega_m$=0.27) are adopted throughout the paper.
\section{Sample presentation and analysis}
\label{sect:sample_and_analysis}
The \lx-\lha\ relation is a useful diagnostic of the accretion physics and radiative output of AGN. Broad \ha\ line emission in AGN mostly originates from gas in the BLR that is photoionised by Lyman continuum photons from the accretion disc, whereas the accretion flow generally produces a hot corona responsible for the X-ray continuum \citep[e.g.][]{1980A&A....86..121S,1993ApJ...413..507H}. 

\begin{figure}
\centering
\includegraphics[width=\columnwidth]{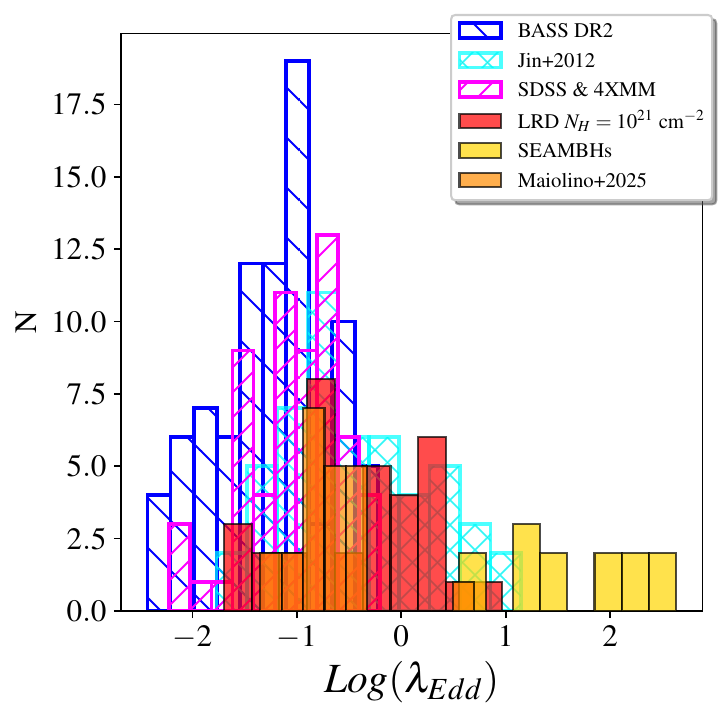}
\caption{\footnotesize Histogram of the distribution of the Eddington ratio values ($\lambda_{\rm Edd}$) of the samples considered in this work. We report: low-redshift type~I AGN from BASS \citep[][blue]{Gupta2024}, low-redshift type~I AGN from SDSSDR16 \citep[][magenta]{Wu_2022}, NLS1 \citep[][cyan]{Jin2012a}, LRDs \citep[][red]{Yue_2024}, JWST-selected AGN \citep[][orange]{Maiolino2025}, and SEAMBHs \citep[][yellow]{Du2014,Du2015}.}
\label{fig:hysto}
\end{figure}

The X-ray emission in AGN mainly originates from a hot corona of relativistic electrons, located in the vicinity of the black hole. Thermal UV/optical photons emitted from the accretion disc are inverse-Compton scattered by the hot electrons into the X-rays, creating a power-law continuum \citep[e.g.][]{1980A&A....86..121S,1993ApJ...413..507H}. Due to the coupling between accretion disc and AGN corona, the relative strength of the X-ray and optical/UV emission is expected to depend on the accretion state, and in particular on the Eddington ratio $\lambda_{\rm Edd}=L_{\rm bol}/L_{\rm Edd}$, \citep[e.g.][]{2021ApJ...910..103L}. Therefore, an empirical correlation between the BLR recombination radiation and the hard X-ray emission is expected, and it has already been explored in samples of low-redshift type~I AGN \citep[e.g.][]{Ho2001,Jin2012b}. Deviations from this empirical relation can signal changes in the structure or energetics of the central engine. In particular, significant suppression of \lx\ relative to \lha\ may indicate coronal inefficiency, obscuration, or fundamental differences in the accretion flow properties, such as those expected in super-Eddington regimes. This is why we decided to explore the \lx-\lha\ relation and the \lx/\lha-$\lambda_{\rm Edd}$ relation in a sample of LRDs and compare it, for the first time, with local AGN over a wide range of accretion rates (see Fig.\,\ref{fig:hysto}). Comparing the position of LRDs in the \lx-\lha\ plane (even through upper limits) to local AGN samples can help interpret their weak X-ray emission, which has implications for the geometry and cooling of the X-ray corona and the role of outflows. Exploring this relation is thus essential to interpreting the growth and radiative efficiency of SMBHs in the early Universe and to testing whether standard accretion models remain valid at high $z$.

The sample considered in this work includes the following objects:
\begin{itemize}
    \item 87 low-redshift ($0.02<z<0.29$, $z_{\rm med}=0.035$), low accreting ($0.003<\lambda_{\rm Edd}<0.6$, $\lambda_{\rm Edd, med}=0.06$), radio-quiet (to minimise contamination of the 2–10 keV emission by jet-related non-thermal processes), X-ray unobscured ($N_H < 10^{21}\,\rm cm^{-2}$) type~I AGN from the BASS sample. 
    We took the broad \lha\ from \citet{2022ApJS..261....5M} and the \lx\ and $\lambda_{\rm Edd}$ from \citet{Ricci2017} and \citet{Gupta2024}, respectively.\\
    \item 61 low-redshift ($0.019<z<0.4$, $z_{\rm med}=0.123$) AGN, optically classified as type~I based on the presence of a broad \ha\ component in SDSS DR16 spectra and cross-matched with the 4th \textit{XMM} Serendipitous Source Catalog (4XMM-DR13). The matching radius is $6''$ for SDSS.\\
    \item 51 low-redshift ($0.03<z<0.37$, $z_{\rm med}=0.16$) X-ray or optically selected sample of NLS1s from \citet{Jin2012a}.\\
    \item 13 low-redshift ($0.01<z<0.18$, $z_{\rm med}=0.04$) NLS1s  belonging to the SEAMBH sample and identified as super-Eddington accreting AGN, $\lambda_{\rm Edd}>1$, through dedicated reverberation-mapping (RM) campaigns \citep{Du2014,Du2015}. For these sources, black hole masses were obtained from RM, while bolometric luminosities were derived from detailed optical-to-X-ray SED fitting  performed by \citet{2016MNRAS.458.1839C}. The X-ray properties used in this work are from \citet{Tortosa2023}.\\
    \item 34 spectroscopically confirmed LRDs ($3.1<z<6.9$, $z_{\rm med}=5.28$) that show broad \ha\ emission lines and no broad forbidden lines (i.e. [OIII]) with the same width as \ha (pointing to a BLR origin for the broad \ha\ rather than to an outflow), from \citet{Yue_2024} and references therein.\\
     \item 22 broad-line type~I AGN ($4.1<z<10.6$, $z_{\rm med}=5.23$) discovered by JWST, whose AGN nature is established through the detection of broad \ha\ emission in JWST/NIRSpec spectra, without any broad forbidden lines, \citep{Maiolino2025}. X-ray constraints for these sources were obtained from deep archival Chandra observations in the GOODS fields, specifically the 2\,Ms Chandra Deep Field–North survey \citep{2003AJ....126..539A} and the 7 Ms Chandra Deep Field–South survey \citep{2017ApJS..228....2L}, as compiled and analysed by \citet{Maiolino2025}.\\
     \item 2 low-redshift ($z=0.10$ and $z=0.16$) LRD analogues recently presented by \citet{Lin2026}.
    \end{itemize}
The main properties and selection criteria of the AGN subsamples considered here are summarised in Table~\ref{tab:samples}. The distribution of the Eddington ratios of the various samples considered in this work are shown in Fig.\,\ref{fig:hysto}.
\begin{table*}
\caption{Summary of the AGN subsamples used in this work, including selection criteria, sample size, and the method adopted to derive key physical parameters.}
\centering
\begin{tabular}{l c l c l l}
\hline\hline
Subsample & $z$ range & Selection & $N_{\rm src}$ & $L_{\rm bol}$ & Ref. \\
\hline
BASS  & $0.02<z<0.29$ & Broad H$\alpha$ + X-ray detected; radio-quiet, $N_H < 10^{21}\,\rm cm^{-2}$ & 87 & SED-based & (1,2,3)\\
SDSS--4XMM & $0.019<z<0.4$ & Broad H$\alpha$ + X-ray detected; radio-quiet, $N_H < 10^{21}\,\rm cm^{-2}$ & 61 & H$\alpha$-based & (4,5) \\
NLS1s & $0.03<z<0.37$ & Optical NLS1 + X-ray detected, $N_H < 10^{21}\,\rm cm^{-2}$ & 51 & SED-based & (6) \\
SEAMBHs & $0.01<z<0.18$ & RM optical NLS1s, $\lambda_{\rm Edd} \gtrsim 1$ & 13 & SED-based & (7,8,9,10) \\
LRDs & $3.1<z<6.9$ & Broad H$\alpha$ detected with JWST/NIRSpec & 34 & H$\alpha$-based & (11)\\
JWST--AGN & $4.1<z<10.6$ & Broad H$\alpha$ detected with JWST/NIRSpec &22 &H$\alpha$-based & (12)\\
low $z$ LRDs & $z \lesssim 0.1$ & Compact, red AGN with broad H$\alpha$ & 2 & H$\alpha$-based & (13)\\
\hline
\end{tabular}
\tablebib{(1)~\citet{2022ApJS..261....5M};(2)~\citet{Ricci2017}; (3)~\citet{Gupta2024}; (4)~\citet{Wu_2022}; (5)~\citet{2020A&A...641A.136W}; (6)~\citet{Jin2012b}; (7)~\citet{Du2014}; (8)~\citet{Du2015}; (9)~\citet{2016MNRAS.458.1839C}; (10)~\citet{Tortosa2023}; (11)~\citet{Yue_2024}; (12)~\citet{Maiolino2025}; (13)~\citet{Lin2026}.}
\label{tab:samples}
\end{table*}
We note that in low-redshift comparison samples (BASS, SDSS–4XMM, NLS1s, and SEAMBHs), all sources have both broad \ha\ and X-ray measurements as a natural consequence of the sample construction and survey depths, rather than as a result of additional selection cuts applied in this work. In particular, the SDSS–4XMM sources are optically classified as type~I AGN based on the presence of a broad \ha\ line component in their SDSS spectra, with no similarly broad [OIII] counterpart, while their X-ray detections arise from archival \xmm observations of sufficient depth at low $z$. Similarly, BASS AGN are bright, nearby systems with well-characterised optical broad-line emission and X-ray detections. In contrast, the LRDs and JWST-selected AGN are primarily selected via rest-frame UV and optical spectroscopy, and their X-ray measurements are almost exclusively limited to upper limits due to their observed X-ray weakness and current sensitivity constraints. Indeed, if their spectra are intrinsically steep (i.e., very soft), then at high $z$ much of the emission is redshifted out of the observed band, leading to a substantial loss of detected photons with current X-ray instruments. This difference reflects observational limitations rather than intrinsic selection biases and should be borne in mind when comparing the distributions.

The treatment of dust attenuation follows the original measurements in the literature and is therefore not fully homogeneous across sub-samples. For the local SDSS type~I AGN, as well as for the BASS AGN sample and the NLS1 sample, the spectra are corrected for foreground Galactic extinction, while no uniform correction for intrinsic (host or BLR) reddening of the broad \ha\ emission is applied. For JWST-selected AGN and LRDs, broad \ha\ luminosities are reported after extinction correction.

Moreover, the bolometric luminosity is not computed uniformly across all sub-samples: it is derived from SED fitting for the SEAMBH sample \citep{2016MNRAS.458.1839C}\footnote{We note that for the SEAMBHs AGN, also \citet{2021ApJ...910..103L} derived SED-based $L_{\rm bol}$. We adopt the values of \citet{2016MNRAS.458.1839C} as they performed detailed SED fitting with the \citet{2012MNRAS.426..656S} code, comparing the observed SEDs with a wide range of accretion disc models spanning different masses, accretion rates, and spins, and accounting for both intrinsic reddening and host-galaxy contamination. By contrast, \citet{2021ApJ...910..103L} derived SED-based $L_{\rm bol}$ for SEAMBHs using the more simplistic templates of \citet{Krawczyk_2013}, resulting in different values for some of the SEAMBHs sources. A more comprehensive SED analysis is underway and will be presented in a forthcoming paper (Kallova et al., in prep.).} as well as for BASS sources \citep{Gupta2024}, whereas in LRDs, JWST-AGN and low-redshift LRD analogues it was estimated from extinction-corrected broad component of the \ha\ line luminosity, adopting the scaling relation provided by \citet{Stern_2012} (i.e.\ $L_{\rm bol}\!\simeq\!130\,L_{{\rm H}\alpha}$), following the same approach applied on the JWST-selected AGN from \citet{Maiolino2025}. Ideally, a full reanalysis using homogeneous methods would be desirable, but this is beyond the scope of the present work. We therefore acknowledge this as a source of systematic uncertainty when comparing different AGN populations.

We also caution that the Eddington ratios of LRDs, JWST-AGN and low $z$ LRDs reported here are derived using $L_{\rm bol}$ obtained from the relation provided by \citet{Stern_2012}. Formally, these values are not lower limits. However, some considerations suggest that the Eddington ratio may be underestimated. First, the $M_{\rm BH}$ of LRDs could be systematically overestimated. As argued by \citet{King2024}, the use of virial relations calibrated on thin-disc AGN may be inappropriate for super-Eddington systems, where the inflated inner disc can shield the BLR and modify the observed line widths, biasing higher values of $M_{\rm BH}$. Second, as shown in \S\,\ref{sect:bolometric}, for SEAMBHs, bolometric luminosities inferred from \lha\ tend to underestimate those obtained from full SED fitting. These effects imply that the quoted $\lambda_{\rm Edd}$ values for LRDs should be regarded as conservative estimates, and may in fact represent lower limits.

Besides comparing the \lx-\lha\ and the \lx/\lha-$\lambda_{\rm Edd}$ relations between LRDs, JWST-selected AGN and local AGN, we checked the relation between the \lx/\lha and $\lambda_{\rm Edd}$ in our entire comparison sample of local AGN to establish whether this ratio depends systematically on the mass-normalised accretion rate, providing a critical benchmark for interpreting the suppressed X-ray emission observed in LRDs and JWST-selected AGN and evaluating whether their deviations from standard scaling relations reflect intrinsic differences or selection effects. We fitted a linear model to the data using the following fitting relation:
\begin{equation}
  \log(L_{2-10\,keV}/L_{\rm H\alpha})=\rm{A}+\rm{B}\log(\lambda_{\rm Edd})
    \label{eq:fitting_rel}
\end{equation}
We checked the possible relation also in two sub-samples, one including the sources with $\lambda_{\rm Edd}\leq 1$ and one including the sources in  which $\lambda_{\rm Edd}> 1$. 

To account for the presence of upper limits in the analysis, we adopted a Monte Carlo–based 'censored fitting' (CF) approach following \citealt{Guainazzi2006,Bianchi2009}. For each fit, we generated $10^4$ Monte Carlo realistions of the dataset. For sources with X-ray detections, values were drawn from Gaussian distributions centred on the measured quantities with widths given by their statistical uncertainties. For sources with X-ray upper limits, values were drawn from a uniform distribution between zero and the measured upper limit. The fitting procedure was applied to each realistion using the \texttt{linmix} Bayesian regression code \citep{Kelly_2007}, and the final best-fit parameters were obtained from the resulting posterior distributions. The values shown in the figures correspond to the nominal observed measurements and upper limits, while the Monte Carlo realistions are used solely to estimate the underlying trends and their uncertainties.

We also investigated the relation between the X-ray bolometric correction, $\kappa_{\mathrm{bol,X}} = L_{\rm bol}/$\lx, and the bolometric luminosity. This diagnostic provides a measure of the relative contribution of the X-ray corona to the total AGN energy budget, and has been widely studied in the literature \citep[e.g.][]{Lusso2010,Netzer2019,Duras2020,Gupta2024}. For some of the samples analysed here $L_{\rm bol}$ is inferred from the broad \ha\ luminosity using the \citet{Stern_2012} relation. In this case, $\kappa_{\rm bol,X}$ is effectively a scaled version of the $L_{\mathrm{H\alpha}}/L_{2-10\,\mathrm{keV}}$ ratio. The comparison with literature values based on SED-derived $L_{\rm bol}$ provides also a useful consistency check and highlights possible systematics in H$\alpha$-based bolometric estimates.
\section{Results and discussion}
\label{sect:Discussion}
\begin{figure*}
\centering
\includegraphics[width=0.49\textwidth]{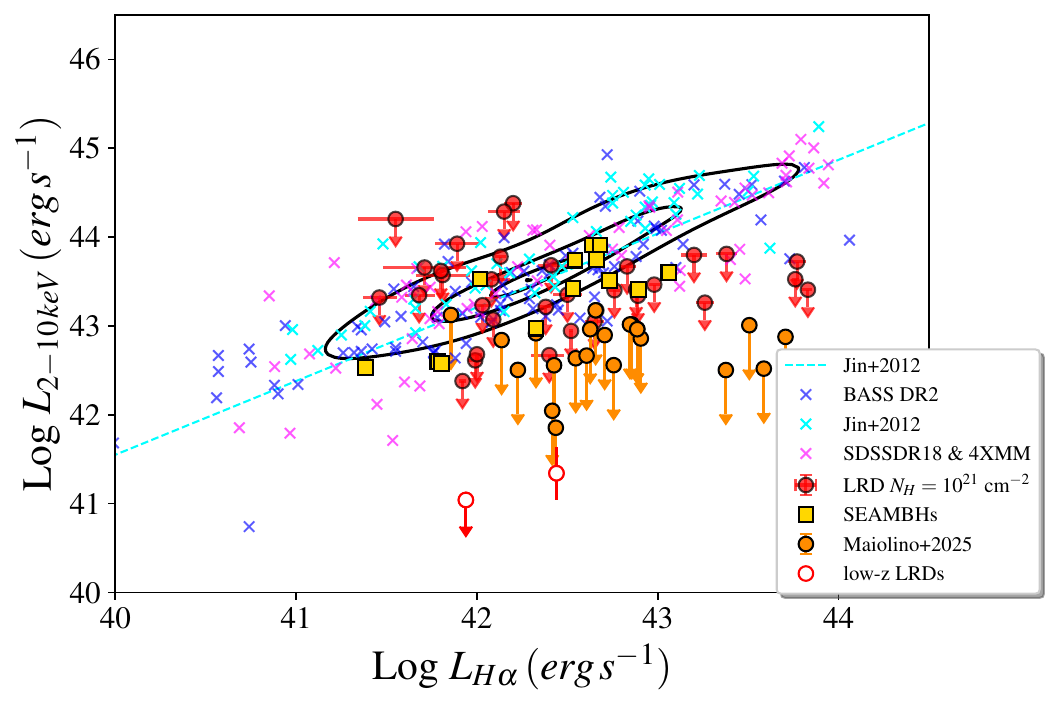}
\includegraphics[width=0.47\textwidth]{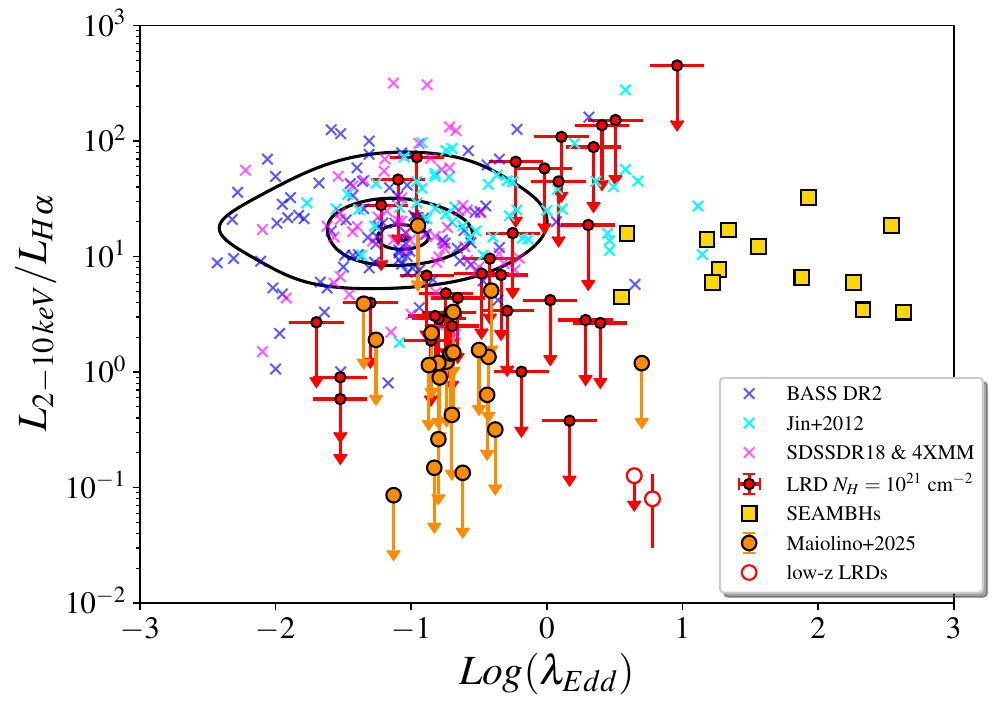}
\caption{\footnotesize {\it Left}: \lx-\lha\ relation; {\it Right}: \lx/\lha-$\lambda_{\rm Edd}$ relation. We report: the BASS sample  \citep[blue crosses,][]{Gupta2024}, the SDSSDR16--4XMM sample \citep[magenta crosses,][]{Wu_2022} and NLS1 sample \citep[cyan crosses,][]{Jin2012a}. Red circles are the upper limits of LRDs from \citet{Yue_2024} derived assuming $N_{\rm H}=10^{21}\,\rm cm^{-2}$. Orange circles are the upper limits of JWST-selected AGN from \citet{Maiolino2025}. Yellow squares are the SEAMBHs from \citet{Tortosa2023}. The dashed cyan line marks the best fit linear relation from \citet{Jin2012a} respectively. White circles with red edges are the low-z LRDs analogues from \citealt{Lin2026}. We report also the $1-2-3\sigma$ contour levels (relative to the peak) for the low-redshift type~I AGN.}
\label{fig:lx_lha}
\end{figure*}
\subsection{\lx-\lha\ plane}
Our analysis on the position of the sources of our sample in the \lx-\lha\ plane (see left panel of Fig.\,\ref{fig:lx_lha}) reveals that the upper limits on X-ray luminosity for LRDs and JWST-selected AGN place them systematically below the \lx–\lha\ relations defined by local type~I AGN. Most of them fall within or near the lower envelope of the SEAMBH distribution, suggesting some level of consistency. 

However, while some LRDs and JWST-selected AGN may share properties with SEAMBHs, such as high accretion rates and suppressed X-ray emission, others appear to lie in a more extreme regime where the X-ray suppression is stronger or the coronal emission is absent. This may imply that they could not completely represent a direct high-redshift analog of the SEAMBH population, but a more extreme evolutionary stage. The presence of broad \ha\ line emission and high inferred $\lambda_{\rm Edd}$ values supports a rapid accretion scenario, yet their weak X-ray emission could reflect either intrinsic changes in the disc--corona system (e.g. altered coronal geometry or radiative coupling) or substantial obscuration. In particular, heavy absorption, potentially reaching the Compton-thick regime, could strongly suppress the observed 2--10\,keV emission even in the presence of an active BLR, especially if the obscuring medium is compact, clumpy, or associated with dense inflows/outflows at early cosmic times \citep{Bianchi2012,2018ARA&A..56..625H,COmastri2025}.

The low $z$ LRDs analogues appear somewhat isolated from most of the local populations. This separation may reflect the intrinsic scatter of this small analogue sample, but it could also be related to the fact that for high $z$ LRDs only upper limits on the X-ray luminosity are available. In this sense, the low $z$ analogues may highlight the parameter space where LRDs would lie if their intrinsic X-ray emission were more firmly constrained.

\subsection{\lx/\lha\ - $\lambda_{\rm Edd}$ plane}
The distribution of sources in the \lx/\lha\ versus $\lambda_{\rm Edd}$ plane reveals distinct trends across AGN populations. Local low-redshift AGN are concentrated along a well-defined locus, as highlighted by the $1\sigma$, $2\sigma$, and $3\sigma$ density contours (see right panel of Fig.\,\ref{fig:lx_lha}). AGN from the BASS survey, which typically exhibit low $\lambda_{\mathrm{Edd}}$, show comparatively high X-ray-to-H$\alpha$ ratios, consistent with a luminous corona contributing significantly to the total emission. At intermediate Eddington ratios, sources such as NLS1s and type~I AGN from the SDSS--XMM cross-match demonstrate a decline in this ratio, indicating a decreasing relative contribution from the X-ray emitting corona, in agreement with the general idea that moving towards higher $\lambda_{\rm Edd}$ the X-ray emission weakens  \citep[e.g.][]{10.1093/mnras/stt920,2021ApJ...910..103L,Laurenti2022,Tortosa2023}. In contrast, SEAMBHs occupy a clearly offset region, characterised by high Eddington ratios and markedly suppressed X-ray emission relative to their H$\alpha$ luminosities. Systematic uncertainties in different black hole mass estimates may contribute to the observed offsets in $\lambda_{\rm Edd}$ among different samples. In particular, BASS, SDSS, and NLS1 samples mostly rely on earlier empirical radius–luminosity relations. Single-epoch virial mass estimators are typically affected by systematic uncertainties of $\sim 0.3-0.5$ dex, depending on the adopted radius–luminosity relation, line-width measurement, and calibration \citep[e.g.][]{2006ApJ...641..689V,2013BASI...41...61S,2018MNRAS.478.1929M}. SEAMBHs are typically characterised using reverberation-based scales corrected for Fe\,{\sc ii} emission. The use of iron-corrected radius–luminosity relations for highly accreting sources \citep{Du_Wang2019} can shift $M_{\rm BH}$ estimates by up to $0.5$ dex relative to standard calibrations. Moreover, recent works have suggested that the observed broad Balmer profiles in some LRDs may not trace purely virialised BLR kinematics, but could be broadened in part by electron scattering in a dense ionised medium, which would bias single-epoch virial masses \citep[e.g.][]{2025arXiv250316595R,Inayoshi25,2025arXiv251000103T}. Other scenarios invoke resonant/scattering effects in Balmer lines in gas-enshrouded ('black-hole star') models \cite[e.g.][]{Naidu25}. At the same time, the dominant role of scattering in setting the line widths is still debated, with recent analyses arguing against a simple, ubiquitous electron-scattering interpretation in at least some well-studied cases \citep[e.g.][]{10.1093/mnrasl/slaf116}. Such systematics would propagate directly into comparable uncertainties in $\lambda_{\rm Edd}$. However, the separation observed between the bulk of sub-Eddington AGN and the SEAMBHs population typically exceeds $\sim1$ dex in $\lambda_{\rm Edd}$, indicating that plausible uncertainties in $M_{\rm BH}$ alone are insufficient to fully account for the observed offset. This suggests that, while mass-estimation systematics may contribute, an intrinsic difference in accretion regime, likely involving slim or radiation-pressure-dominated discs \citep{2019Univ....5..131C}, is still required to explain the location of SEAMBHs, and possibly LRDs, in the high-$\lambda_{\rm Edd}$ parameter space.

At high accretion rates, especially at super-Eddington regimes, the X-ray spectrum tends to steepen as the corona experiences stronger Compton cooling from the enhanced photon density of the accretion disc, associated with large $\lambda_{\rm Edd}$ \citep{Wang2004}. In addition, pair production at high compactness may further regulate the coronal temperature, reinforcing the observed trend towards softer spectra at high $\lambda_{\rm Edd}$ \citep[e.g.][]{Fabian2017,Kara_2017,Ricci_2018,Tortosa22,Tortosa2023}. Indeed, optically thick geometrically thin accretion discs have long been used as a baseline framework to interpret several aspects of the observed spectral energy distribution (SED) of AGN with moderate Eddington ratios ($\lambda_{\rm Edd}\in[0.01;0.3]$, \citealt{Koratkar1999, Capellupo2015}) although their applicability remains debated \citep{Blaes2007,Davis2011,Jin2012b,Done2012,Laor2014,Antonucci2023}. At higher accretion rates, these models predict a transition to geometrically thick (i.e., slim disc, \citealt{2019Univ....5..131C}), and the nature of the accretion flow is expected to change dramatically by photon trapping through electron scattering in dense matter and advection cooling. Moreover, strong gas outflows are naturally expected during super-Eddington accretion episodes \citep{Ballantyne2011,Zubovas2012,Jiang24} due to the intense radiation pressure associated with these events. The presence of outflowing disc winds has also been observed in some low $z$ \citep{Jin2017,Giustini2019,Tortosa22} and high $z$ QSOs \citep{Chartas03, Lanzuisi2012, Vignali2015, Lanzuisi2016, Tortosa2024} accreting close to the Eddington limit and in ultraluminous X-ray sources \citep[ULXs,][]{2017AN....338..234P}. 

Interestingly, LRDs and JWST-selected AGN - although only have upper limits in X-ray luminosity - tend to cluster close to the tail of the local AGN distribution of \lx/\lha\ plane, potentially overlapping with the region occupied by NLS1s, which are also known to accrete close to or above the Eddington limit.  Alternatively, some LRDs may lie within the region occupied by SEAMBHs, and in several cases even below it. While the upper-limit nature of these measurements prevents a definitive placement, the data suggest that LRDs may share similar accretion regimes with SEAMBHs despite residing at much higher redshifts. Also the low $z$ LRDs analogues seem to follow the general trend of decreasing ratio between X-ray luminosity and \ha\ luminosity with increasing Eddington ratio. However, given the very limited sampling (only two sources), and the fact that one of them is constrained only by an upper limit on \lx, no firm conclusions can be drawn.

The observed X-ray weakness of LRDs may reflect intrinsic changes in the accretion flow structure at high Eddington ratios. In the super-Eddington regime, the accretion disc becomes geometrically and optically thick, with strong radiation pressure capable of altering or even suppressing the formation of a stable X-ray emitting corona \citep[e.g.][]{Jiang2019, 2020ARA&A..58...27I}. One possibility is that enhanced Compton cooling by the high photon density of the accretion disc lowers the electron temperature in the corona, leading to a steepening or suppression of the X-ray continuum. Additionally, powerful radiation-driven winds, expected in super-Eddington systems \citep{Jiang2019,Okuda_2021,Zhang2024}, may disrupt the vertical stratification or magneto-thermal structure necessary to sustain a hot corona. Such disruption may also be relevant to scenarios in which jet launching fails, as in the ‘aborted jet’ model \citep{Ghisellini2009}, where outflows initiated near the black hole lose collimation and dissipate their energy close to the accretion flow rather than escaping. A similar mechanism may underlie the intermittent disappearance of the X-ray corona observed in Mrk\,335 \citep{Gallo2018}, where the hot plasma seems to collapse or be quenched. In both cases, strong winds or radiation pressure could destabilise vertical structures above the disc, preventing the long-term survival of either a compact corona or a nascent jet. Coronal quenching by powerful radiation-driven outflows would be consistent with the observed X-ray weakness of low-redshift SEAMBHs \citep{Jin2012a,Tortosa22,Tortosa2023}, which also exhibit low \lx/\lha\ and steep X-ray spectra. Therefore, the X-ray weakness of LRDs may be a physical consequence of high accretion rates, rather than solely due to line-of-sight obscuration. 

However, given that the vast majority of LRDs and JWST-selected AGN are currently constrained only by X-ray upper limits, the available data do not allow us to discriminate between obscuration-dominated and accretion-driven scenarios. Both possibilities remain viable, and deeper X-ray observations are required to determine whether the suppressed X-ray emission arises from heavy nuclear obscuration, intrinsic coronal weakness, or a combination of the two. In this context, recent work by \citealt{2025ApJ...989L..30S} reports no significant X-ray detection by stacking X-ray observations of 55 LRDs in the Chandra Deep Fields, and argue against a population of unobscured, intrinsically luminous super-Eddington accretors. This result further highlights the degeneracy between intrinsic X-ray weakness and heavy nuclear obscuration, and emphasises that current X-ray data alone are insufficient to uniquely determine the dominant physical mechanism responsible for the suppressed X-ray emission in LRDs.
\subsection{\lx/\lha\ versus $\lambda_{\rm Edd}$ relation}
\begin{figure}
\centering
\includegraphics[width=\columnwidth]{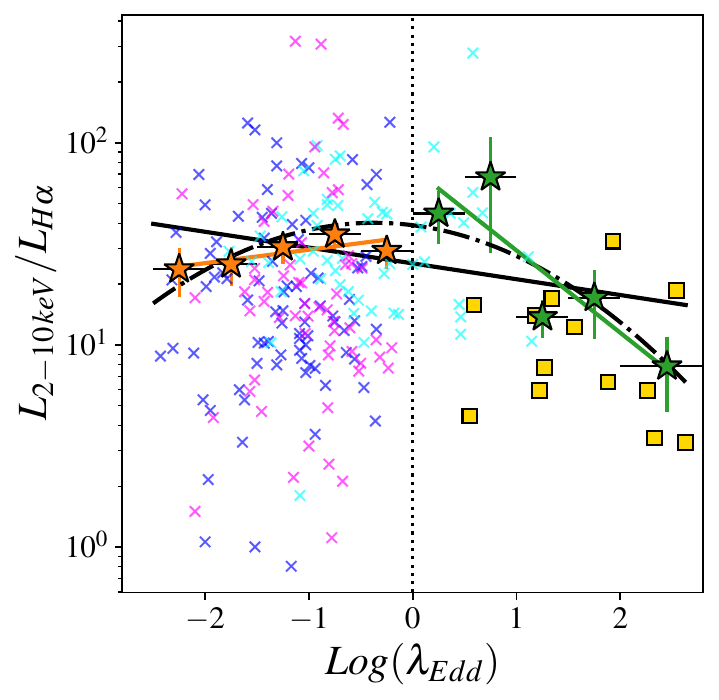}
\caption{\footnotesize \lx/\lha-$\lambda_{\rm Edd}$ relation. We used the same colour code as Fig.\,\ref{fig:lx_lha}. Solid black line is the best fitting linear relation considering the complete sample. We reported also the best fitting linear relation for the sub- (orange) and super- (green) Eddington sources. The stars represent the mean value in each $\lambda_{\rm Edd}$ bin and are shown for illustrative purposes only. Given the large intrinsic scatter (spanning $\geq 2$ orders of magnitude in \lx/\ha), these averages are not used for $\chi^2$ minimisation or quantitative inference. A quadratic fit (dotted-dashed curve) is shown to illustrate curvature in the trend. This figure includes only low-redshift AGN with joint detections in both X-rays and \ha; no upper limits are included in this analysis.
}
\label{fig:fit}
\end{figure}
To our knowledge, a systematic analysis of the relation between \lx/\lha\ and $\lambda_{\rm Edd}$ has not been previously conducted in the literature, although both quantities have been extensively studied separately in various AGN samples (e.g. \citealt{Ho2001,Panessa2006,Ho2008,Ho2009,Jin2012a,Du2015,Wang2014}). Thus, we checked the presence of this relation (see Fig.\,\ref{fig:fit}). Our analysis reveals a tentative anti-correlation between the X-ray to H$\alpha$ luminosity ratio and the Eddington ratio in our sample of local AGN. When considering the full sample, we find a Spearman correlation coefficient of $\rho = -0.43$ with a statistical significance of approximately $1.6\sigma$ (corresponding to $P_{\rm value}=0.109$). This suggests a weak-to-moderate trend. The trend becomes notably stronger within the super-Eddington subsample, where the correlation coefficient reaches $\rho = -0.90$ with a statistical significance of $\sim 2.5\sigma$ (corresponding to $P_{\rm value}=0.012$). This indicates a more substantial anti-correlation: as $\lambda_{\rm Edd}$ increases, the relative X-ray output (compared to broad H$\alpha$ emission) decreases. This behaviour is consistent with theoretical expectations for slim accretion discs and super-Eddington flows, where X-ray emission may be suppressed due to photon trapping, coronal evaporation, or increased mass-loading in radiation-driven winds. The suppressed \lx\ may also reflect changes in the disc-corona geometry at high accretion rates, as in slim-discs photon trapping and advection reduce the efficiency of radiative escape, while the geometrically thick inner disc can shield parts of the corona and BLR, together reinforcing the tendency toward intrinsically softer X-ray continua. In contrast, the sub-Eddington AGN does not show a significant correlation between \lx/\lha\ and $\lambda_{\rm Edd}$ ($\rho = 0.60$, with statistical significance $\sim 1.1\sigma$ corresponding to $P_{\rm value}=0.274$). This result may reflect either intrinsic X-ray variability, variability in the line emission, or a genuine lack of a systematic trend in this accretion regime. The absence of a clear relation at low $\lambda_{\mathrm{Edd}}$ is consistent with observations for standard accretion flows, where both X-ray and optical emission scale more uniformly with bolometric output.

To explore possible departures from a single power-law behaviour, we also fitted a second-order polynomial to the \lx/\lha-$\lambda_{\rm Edd}$ relation (see Fig.~\ref{fig:fit}). The curvature captured by this fit suggests that the ratio remains comparatively flat or only mildly varying at low accretion rates, while it declines more steeply as $\lambda_{\mathrm{Edd}}$ approaches and exceeds unity. The vertex of the quadratic relation is located at $(\log \lambda_{\rm Edd}$, log \lx/\lha) $\simeq (-0.37,\,1.60)$, corresponding to $\lambda_{\rm Edd} \sim 0.4$ and \lx/\lha\ $\sim 40$, implying that the relative luminosity of the corona with respect to the broad H$\alpha$ emission reaches a maximum around moderate accretion rates. Physically, this behaviour is consistent with a transition from radiatively efficient, geometrically thin discs (with an energetically significant corona) to slim-disc regimes in which enhanced Compton cooling, photon trapping, and radiation-driven winds suppress the coronal hard X-ray output and/or modify the BLR response. The observed curvature may therefore be interpreted as the imprint of two distinct accretion regimes acting at opposite ends of the sequence: at low $\lambda_{\mathrm{Edd}}$, radiatively inefficient flows (RIAF/ADAF-like) reduce the seed-photon supply and suppress hard X-ray output \citep{narayan1994,Yuan2014}, while at high $\lambda_{\mathrm{Edd}}$, slim-disc physics leads to photon trapping and strong radiation-driven outflows that further quench the corona \citep{Sadowski2016,Jiang2019,2020ARA&A..58...27I}.

We emphasise that the polynomial fit is used only as an empirical description of the observed trend; it has no physical basis and may be affected by the presence of upper limits and by the mixing of different source populations. It is purely phenomenological and is used only as a guide to the eye. They are not intended to provide a statistically rigorous description of the data across heterogeneous AGN populations, nor to identify a sharp transition across the Eddington limit.

Overall, these findings suggest that the \lx/\lha\ ratio is not constant, but likely depends on the accretion state. In particular, super-Eddington AGN exhibit systematically lower X-ray emission relative to their broad-line luminosities. This trend is particularly relevant when evaluating high-redshift AGN candidates such as LRDs, which also show suppressed X-ray emission. The observed suppression of \lx\ at high $\lambda_{\mathrm{Edd}}$ is consistent with previous results linking X-ray weakness to enhanced accretion, such as the $\alpha_{\mathrm{OX}}$–$\lambda_{\mathrm{Edd}}$ relation \citep[e.g.][]{Shemmer2008,Lusso2016,2021ApJ...910..103L,Lambrides2024}. Recent work further connects this behaviour to highly accreting AGN and changing-look phenomena \citep[e.g.][]{Li2024}. In the context of LRDs, models such as those of \citealt{Liu2025} provide additional evidence that these black holes may indeed be in a highly accreting state. However, \citealt{Liu2025}'s model interpret LRDs within a framework of spherically symmetric super-Eddington accretion onto a central SMBH. This, in principle, may imply different geometries and radiative properties, and should therefore be regarded as a complementary interpretation.
Our results imply that the low \lx/\lha\ values seen in LRDs does not necessarily indicate or unusual geometry, but could instead reflect a physical continuum of accretion properties extending from local super-Eddington sources to the early Universe \citep{Berton2025}. However, given the moderate statistical significance and the relatively small number of high-$\lambda_{\rm Edd}$ sources, these trends should be treated as suggestive rather than conclusive. 
\subsection{$\kappa_{\mathrm{bol,X}}$ versus $L_{\rm bol}$}
\label{sect:bolometric}
\begin{figure*}
\centering
\includegraphics[width=\textwidth]{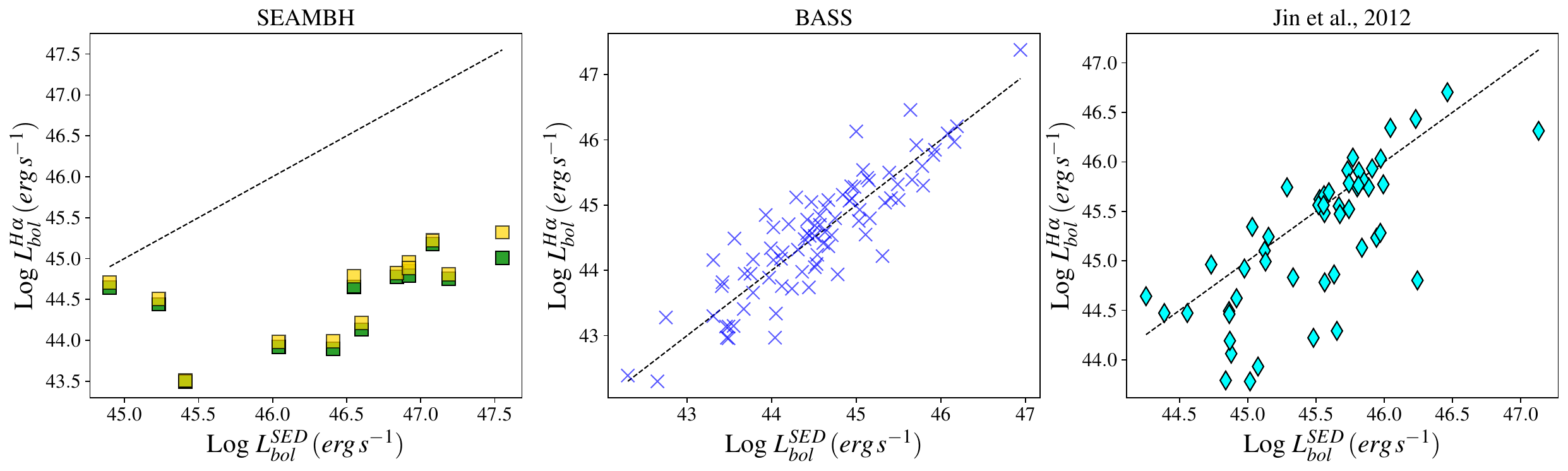}
\caption{\footnotesize Bolometric luminosity estimated from SED-fitting ($L_{\rm bol}^{\rm SED}$) vs bolometric luminosity derived from the broad H$\alpha$ luminosity following \citealt{Stern_2012} ($L_{\rm bol}^{H\alpha}$). \textit{Left panel:} $L_{\rm bol}^{H\alpha}$ when using the observed \lha\ (green squares) and the extinction corrected one (yellow squares) for SEAMBHs. \textit{Central panel}: the BASS AGN (blue crosses). \textit{Right panel}: NLS1s galaxies from \citet{Jin2012a} (cyan diamonds). The dashed black line indicates the one-to-one relation in each panel.}
\label{fig:lbol_comparison}
\end{figure*}
We further examined the relation between the X-ray bolometric correction, $\kappa_{\mathrm{bol,X}} = L_{\rm bol}/L_{2-10\,\rm keV}$, and the bolometric luminosity. This quantity traces the relative weight of the X-ray corona in the overall AGN energy budget, and has been extensively investigated in previous works \citep[e.g.][]{Lusso2010,Netzer2019,Duras2020,Gupta2024}.

Before discussing the relation between $\kappa_{\mathrm{bol,X}}$ and $L_{\rm bol}$, we want to remind that there are some differences between the $L_{\rm bol}$ estimates adopted in this work: for SEAMBHs, the NLS1 from \citealt{Jin2012a} and the BASS AGN, SED-based $L_{\rm bol}$ estimates are available, whereas in LRDs and JWST-selected AGN $L_{\rm bol}$ was estimated from extinction-corrected \lha, adopting the scaling relation provided by \citealt{Stern_2012}. We therefore performed a direct comparison between $L_{\rm bol}$ derived from the broad \ha\ luminosity (using the \citealt{Stern_2012} calibration) and $L_{\rm bol}$ obtained from SED fitting where available. The results of this comparison are shown in Fig.~\ref{fig:lbol_comparison}

Interestingly, the bolometric luminosities inferred from the broad \ha\ emission line using the \citet{Stern_2012} calibration are systematically lower than the SED–based $L_{\rm bol}$ for SEAMBHs (see left panel of Fig.~\ref{fig:lbol_comparison}) while for slowly accreting AGN (e.g. the BASS type~I AGN) there is no systematic underestimation of $L_{\rm bol}$ when using \ha-based estimates (see central panel of Fig.~\ref{fig:lbol_comparison}), indicating that this method is reliable for standard, sub-Eddington accretion regimes. This discrepancy between $L_{\rm bol}$ estimated from \lha\ and the one extrapolated from SED-fitting suggests that the broad \ha\ emission line in SEAMBHs is systematically weaker than expected given their true radiative output. Although previous studies of SEAMBHs did not specifically investigate \ha\ equivalent widths, they consistently reported systematically low H$\beta$ equivalent widths \citep{Du_Wang2019}, indicating similar trends among the Balmer lines. Several effects associated with high accretion rates can account for this. 

We tested whether dust attenuation could explain the lower \ha-based $L_{\rm bol}$ values by correcting \lha\ using the visual extinction value $A_V$ inferred by \citet{2016MNRAS.458.1839C} obtained from thin-disc SED modeling with Small Magellanic Cloud attenuation curve. The $A_V$ values range for SEAMBHs goes from $0.027$ to $0.959$. Then we converted $A_V$ to the extinction at the \ha\ wavelength, and corrected the observed \ha\ luminosities accordingly. We then recomputed the bolometric luminosities using the \citep{Stern_2012} calibration after the extinction correction. Correcting \lha\ for the extinction increases the \ha-based $L_{\rm bol}$ by a median of only $\approx 0.06$ dex ($\approx16\%$, with a maximum of $0.31$ dex). Extinction corrected \ha-based $L_{\rm bol}$ still underestimates the SED-based $L_{\rm bol}$ by a median of $\approx 2.0$ dex (a factor $\approx100$, see left panel of Fig.~\ref{fig:lbol_comparison}), indicating that dust attenuation alone cannot explain the discrepancy and that additional effects are likely required. In super-Eddington (slim–disc) flows the inner geometrically thick funnel produces strong self–shadowing, anisotropically illuminating the BLR and thereby reducing the ionising photon budget that reaches \ha–emitting gas \citep{Wang2014_self-shadowing}. This same high–$\lambda_{\rm Edd}$ population is known to deviate from the canonical BLR radius-luminosity relation, i.e. the 'iron–corrected' ratio-luminosity trend, consistent with non-standard BLR illumination in fast accretors \citep{Du_Wang2019}. Second, high-$\lambda_{\rm Edd}$ sources occupy the extreme end of the quasar main sequence (NLS1–like), where Balmer lines tend to have smaller equivalent widths and strong Fe\,\textsc{ii} emission, again pointing to changes in BLR conditions at high accretion rates \citep[e.g.][]{Marziani2018}. Photoionisation modeling further suggests that the BLR covering factor likely decreases with increasing $\lambda_{\rm Edd}$, which naturally lowers recombination-line output (including \ha) at fixed $L_{\rm bol}$ \citep{Ferland2020}. Together, anisotropic disc illumination (self–shadowing), BLR structural changes, and a reduced covering factor in high–$\lambda_{\rm Edd}$ systems provide a coherent explanation for why \ha-based $L_{\rm bol}$ underestimates the SED–based $L_{\rm bol}$ in SEAMBHs.

\begin{figure*}
\centering
\includegraphics[width=0.9\textwidth]{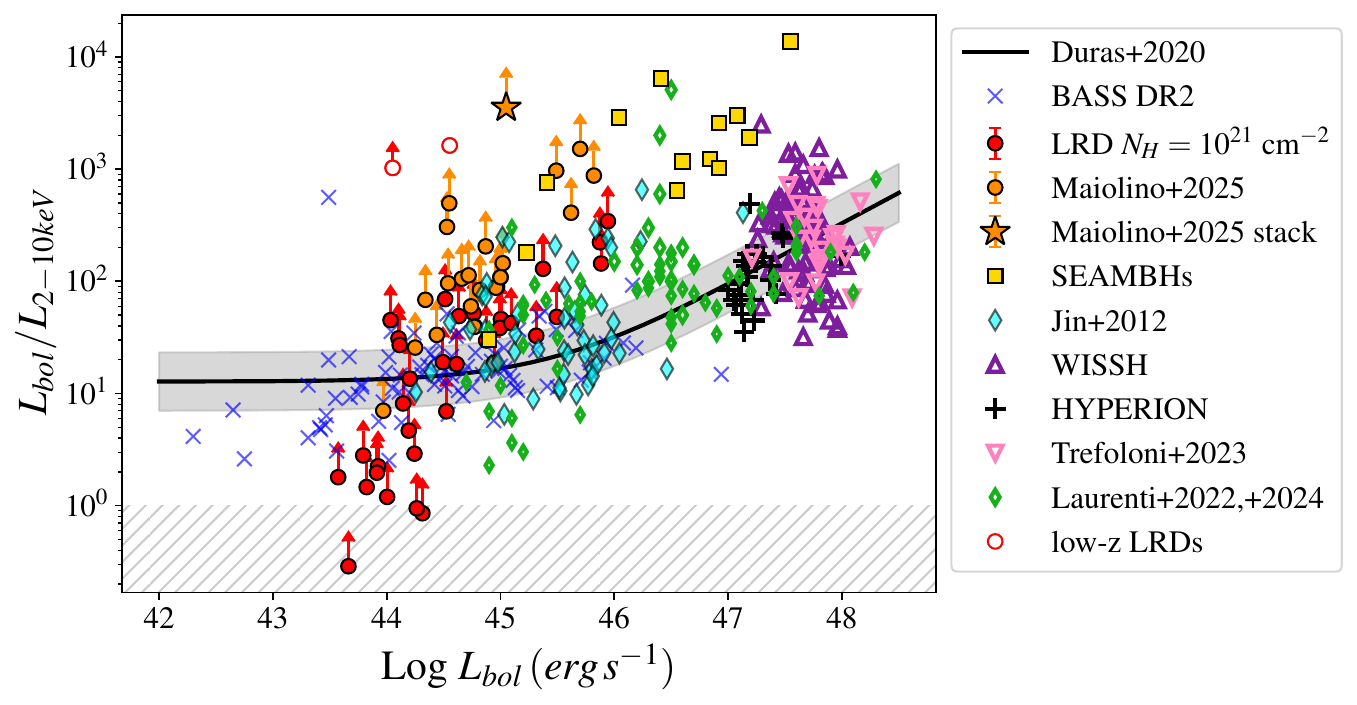}
\caption{\footnotesize X-ray bolometric correction, $\kappa_{\mathrm{bol,X}} = L_{\rm bol}/L_{2-10\,\rm keV}$, as a function of $L_{\rm bol}$. Shown are LRDs \citep[][red circles]{Yue_2024}, JWST-selected AGN \citep[][orange circles]{Maiolino2025}, BASS AGN \citep[][blue crosses]{Gupta2024}, SEAMBHs \citep[][yellow squares]{Tortosa2023}, and NLS1s \citep[][cyan diamonds]{Jin2012a}. The orange star marks the stacked X-ray measurement of the non-detected sources from \citealt{Maiolino2025}. Also shown are low $z$ LRDs analogues \citep[][red empty circles]{Lin2026}, hyper-luminous QSOs at $z\sim 2-4$ from the WISSH sample \citep[][purple empty triangles]{Zappacosta2020,2025A&A...702A.114D}, hyper-luminous QSOs at $z>6$ from the HYPERION sample\citep[black plus markers,][]{Tortosa2024,Saccheo2025}, SDSS DR7 QSOs at $z\simeq 3.0-3.3$ \citep[][pink empty inverted triangle]{Trefoloni2023} and $0.4<z<3.3$ radio-quiet highly accreting AGN \citep[][green diamonds]{Laurenti2022,2024A&A...689A.337L}. $L_{\rm bol}$ for BASS, SEAMBHs and NLS1 is from SED-fitting, while for LRDs and JWST-selected AGN $L_{\rm bol}$ is derived from the broad H$\alpha$ luminosity following \citealt{Stern_2012}, as in \citealt{Maiolino2025}. The black line indicates the relation from \citealt{Duras2020}, with the shaded region showing its $1\sigma$ uncertainty. The hatched region indicates the unphysical regime where $\kappa_{\rm bol,X}<1$.}
\label{fig:kbol_lbol}
\end{figure*}

In Fig.~\ref{fig:kbol_lbol} we report $L_{\rm bol}/L_{2-10\,\rm keV}$ versus $L_{\rm bol}$. The $L_{\rm bol}$ for LRDs and JWST-selected AGN is extrapolated from the \lha\ using the \citet{Stern_2012} scaling relation. $L_{\rm bol}$ for BASS, NLS1s and SEAMBHs samples is the one obtained from SED-fitting reported in the literature. We report in the figure for comparison other samples such as: low $z$ LRDs analogues \citep{Lin2026}, hyper-luminous QSOs at $z\sim 2-4$ from the WISSH sample \citep{Zappacosta2020,2025A&A...702A.114D}, hyper-luminous QSOs at $z>6$ from the HYPERION sample \citep{Zappacosta2023} for which X-ray luminosities are from \citealt{Tortosa2024} while $L_{\rm bol}$ are obtained via SED-fitting from \citealt{Saccheo2025}, SDSS DR7 QSOs at $z\simeq 3.0-3.3$ \citep{Trefoloni2023}, and $0.4<z<3.3$ radio-quiet highly accreting AGN \citep{Laurenti2022,2024A&A...689A.337L}. We also compare the values with the relation from \citep{Duras2020}. The X-ray bolometric corrections of \citep{Duras2020} are calibrated using $L_{\rm bol}$ from full SED fitting, and a direct comparison with \ha-based $L_{\rm bol}$ estimates is therefore not strictly self-consistent. However, for high $z$ LRDs and JWST-selected AGN, reliable SED-based $L_{\rm bol}$ are generally unavailable, making \ha-based estimates the only practical proxy for $L_{\rm bol}$. 
In this figure SEAMBHs populate the high $\kappa_{\rm bol,X}$ (i.e., X-ray–weak) regime, partially overlapping with LRDs and JWST-selected AGN; for the latter, we note that they are lower-limits and may be even more X-ray suppressed \citep{Yue_2024, Maiolino2025}. The JWST-selected AGN provide an important high $z$ comparison sample, with uniformly measured \ha-based $L_{\rm{bol}}$ and deep \textit{Chandra} data. Although most are only constrained via upper limits, their stacked detection indicates systematically high $k_{\rm{bol,X}}$ values, lying two orders of magnitude above the local $L_{\rm{bol}}/$\lx\ relation: their X-ray output is far weaker than expected for their bolometric luminosities, fully consistent with the X-ray weak regime. \citet{Maiolino2025} discuss three main scenarios: (i) Compton-thick, dust-poor absorption, likely by broad line regions clouds with unusually high covering factors, which can suppress X-rays without fully extinguishing broad lines; (ii) intrinsic X-ray weakness (e.g. linked to high accretion rates or inefficient/hot coronae), as seen in NLS1s and SEAMBHs; and (iii) in a minority of cases, misidentification of AGN, though this is deemed unlikely for the bulk of the population. This finding further supports the interpretation that suppressed coronal emission is a generic outcome of high accretion rates at early cosmic times. In addition to absorption, super-Eddington accretion can naturally give rise to a Compton-thick, high-covering-factor envelope surrounding the black hole, which in turn leads to strong attenuation of the X-ray emission \citep[e.g.][]{2016ApJ...822...20N,2017PASJ...69...92K,2019ApJ...871..115Z}. At $\lambda_{\rm Edd}\gtrsim 1$, the inner accretion flow is expected to become geometrically thick and radiation-pressure dominated (i.e.\ a slim-disc regime), favoring the launch of powerful radiation-driven winds from small radii \citep[e.g.][]{2000NewAR..44..435M,2009PASJ...61L...7O}. If the mass outflow rate becomes comparable to the inflow rate, these winds can produce a dense, high-covering-factor envelope around the central engine, with column densities reaching the Compton-thick regime ($N_{\rm H}\gtrsim10^{24}\,\mathrm{cm^{-2}}$) \citep[e.g.][]{2014PASJ...66...48T}. In this scenario, the hard X-ray emission can be strongly attenuated by a combination of photoelectric absorption, Compton scattering, and reprocessing in the optically thick outflow, even if broad-line emission remains detectable along favorable sightlines \citep[e.g.][]{2017A&A...607A..28M}. More recent modeling \citep{Madau2025} further shows that anisotropy and shadowing in thick discs can suppress the observed X-ray emission and high-ionisation lines at high inclinations. 

X-ray weakness at high $\lambda_{\rm Edd}$ is observed in nearby fast accretors and has been linked to episodes of intrinsically weak coronae and/or orientation/absorption effects in slim–disc geometries \citep[e.g.][]{Maithil2024}.  A recent theoretical work also shows that mildly super-Eddington accretion with geometrically thick inner discs naturally yields intrinsically X-ray–weak SEDs, especially away from polar sightlines \citep{Pacucci2024}. Moreover, it has been shown \citep{Zappacosta2020} that X-ray weakness, in hyper-luminous ($L_{\rm bol} \gtrsim 10^{47}$ erg s$^{-1}$) quasars at $z\sim2-4$, strongly correlates with the $\rm C\,\textsc{iv}$ blueshift, a tracer of extreme disc winds. The physics linking disc winds and coronal X-ray emission is further supported by higher-redshift observations. \citet{Tortosa2024} report a significant positive correlation between $\rm C\,\textsc{iv}$ outflow velocity and the steepness of the X-ray spectrum, likely driven by slim-disc geometry and high accretion rates in  the HYPERION \citep{Zappacosta2023} hyper-luminous ($L_{\rm bol}>10^{47}\,erg\,s^{-1}$) QSOs at $z > 6$. The X-ray spectra of HYPERION QSOs are found to be systematically steeper than those of comparable lower-redshift sources. This behaviour is consistent with intrinsically cooler coronae possibly associated with high accretion rates. In the observed energy band, such steep spectra naturally lead to apparent X-ray weakness: for larger photon indices, a larger fraction of the emitted power is redistributed towards lower energies, reducing the flux measured in the rest-frame $\sim2$--$50$ keV band and resulting in lower observed X-ray luminosities.

ASPIRE QSOs at $z>6.5$ \citep{Yang2023} extend the explored parameter space in $M_{\rm BH}$ and optical spectral properties (H$\beta$, [O\,III], Fe\,II) into the reionisation era. Their consistency with low $z$ eigenvector 1 trends and the presence of broad, blueshifted [O\,III] in some cases suggest similarly high accretion states. 

Recent theoretical works suggest that the X-ray weakness of JWST-selected AGN, including LRDs, may be an intrinsic feature of super-Eddington accretion. \citealt{Madau_Haardt2024} proposed that in this regime the hot corona becomes embedded in a funnel-like geometry, leading to efficient Compton cooling and extremely soft, faint X-ray spectra. 

Moreover, our findings are consistent with recent work by \citet{Lambrides2024}, who studied a sample of broad-line AGN at $z \gtrsim 5$ in the deepest {\it Chandra} fields and found no significant X-ray detections, with upper limits placing them well below the expected $\alpha_{\rm OX}$ relation. These sources also lack high-ionisation UV lines (e.g. \ion{C}{IV}, He\,{\sc ii}), even if accounting for dust attenuation, which, together with their strong Balmer emission, were interpreted as signatures of super-Eddington accretion. Radiative transfer modeling shows that slim-disc accretion naturally explains this combination of properties, producing intrinsically weak coronal emission and a softened ionising continuum \citep[see also][]{Wang2014_self-shadowing}. This picture aligns with our finding: SEAMBHs, LRDs, and JWST-selected AGN are all located in the X-ray–weak, high-$\kappa_{\rm bol,X}$ regime. This overlap may be consistent with scenarios in which highly accreting systems exhibit intrinsically weaker X-ray emission due to changes in the disc–corona structure. This overlap may be consistent with scenarios in which highly accreting systems exhibit intrinsically weaker X-ray emission due to changes in the disc–corona structure. However, given that the X-ray luminosities for LRDs and JWST-selected AGN are largely based on upper limits under the assumption of mild obscuration ($N_H = 10^{21} \rm cm^{\rm -2}$), they should not be interpreted as strict constraints on the intrinsic X-ray luminosity. Recent studies have suggested that a significant fraction of these sources may be affected by heavy, and in some cases Compton-thick, obscuration \citep[e.g.][]{Yue_2024,Ananna2024,Maiolino2025,2025A&A...700A..12M,2025ApJ...989L..30S}. In such scenarios, the intrinsic \lx\ could be orders of magnitude higher than the observed limits, which would shift these sources towards lower $\kappa_{\rm bol,X}$ values, potentially consistent with standard AGN relations. Therefore, the high $\kappa_{\rm bol,X}$ values inferred here for LRDs and JWST-selected AGN should be regarded as upper limits, and the apparent overlap with SEAMBHs in $\kappa_{\rm bol,X}$ space may reflect either intrinsic coronal suppression, heavy obscuration, or a combination of both. This interpretation is further complicated by emerging evidence that LRD-like systems may constitute a transitional AGN phase rather than a fundamentally distinct population. \citet{Fu2025} report two LRDs that appear to be evolving into more `classical' quasars, while the `X-ray Dot' presented by \citet{Hviding2026} has been interpreted as either an exotic, dust-powered source or a late-stage analogue of an LRD. Complementarily, \citet{Li2025} identify obscured X-ray AGN at $z\sim3$ that exhibit hot dust emission and strong He\,{\sc i} absorption, highlighting that dense circumnuclear structures and complex radiative-transfer effects may play a critical role in producing the observed X-ray weakness in JWST-selected AGN. The emerging evidence that at least a subset of LRD-like sources may be physically connected to more conventional quasar phases therefore strengthens the interpretation that X-ray weakness could be associated with a transient phase of rapid black-hole growth and/or intense nuclear reprocessing, during which heavy obscuration may naturally arise as part of the evolutionary sequence.

\section{Conclusions}
\label{sect:Conclusions}
We have investigated the nature of Little Red Dots (LRDs), a population of high-redshift ($z \gtrsim 4$) compact sources with broad H$\alpha$ emission and extremely faint X-ray emission, by comparing their properties with those of local AGN and high-z and JWST-selected AGN across a wide range of accretion regimes. In particular, we tested whether LRDs could be considered high-redshift analogues of super-Eddington accreting massive black holes (SEAMBHs), which are known to exhibit similarly low X-ray luminosities relative to their broad-line emission.

Our analysis shows that, given the current X-ray upper limits, LRDs could lie in a region of the \lx-\lha\ plane similar to that occupied by SEAMBHs, or possibly below it (see left panel of Fig.~\ref{fig:lx_lha}). We find a significant anti-correlation between the \lx/\lha\ ratio and Eddington ratio among the super-Eddington subset of our local comparison sample (see Fig.~\ref{fig:fit}). This trend is consistent with a scenario in which coronal emission becomes increasingly suppressed at high accretion rates, possibly due to disc geometrical changes or photon trapping effects, as predicted in slim disc models. The fact that the current X-ray upper limits for LRDs lie in a similar region of \lx/\lha\ vs. $\lambda_{\rm Edd}$ parameter space as the anti-correlation observed in local super-Eddington AGN (see right panel of Fig.~\ref{fig:lx_lha}), suggests that they could be governed by similar accretion physics as SEAMBHs, albeit in more extreme environments. However, given that most LRDs are undetected in X-rays, their true location relative to this trend remains uncertain. Alternatively, the systematic X-ray weakness of LRDs may signal physical conditions unique to the early Universe, including high gas densities, rapid inflows, or substantial obscuration, any of which could alter the structure of the corona and its emission. A mixed scenario is therefore likely, in which some sources are predominantly obscured, others are intrinsically X-ray weak due to super-Eddington accretion, and yet others may be affected by a combination of both effects. In addition, observational constraints need to be taken into account: at high $z$, the combined effects of instrumental sensitivity, limited angular resolution, and the steep X-ray spectra expected for rapidly accreting sources can significantly reduce the detected photon counts, which may in turn bias the measured X-ray luminosities towards smaller values. Deeper, higher-resolution X-ray observations will be essential to disentangling these effects and establishing the dominant drivers of X-ray weakness in high-redshift AGN.

An intriguing result is that the low $z$ LRDs analogues appear even more X-ray weak than SEAMBHs, highlighting that local analogues may probe regimes of suppressed coronal activity beyond those observed in well-studied super-Eddington systems.

We find that bolometric luminosities of SEAMBHs inferred from broad \ha\ are systematically lower than those obtained from SED fitting (see Fig.~\ref{fig:lbol_comparison}), implying that \ha\ may be weaker than expected in sources undergoing fast SMBH accretion. We tested whether dust attenuation could explain the lower $L_{\rm bol}$ \ha-based values by correcting \lha\ for the intrinsic reddening inferred from SED fitting. This correction does not remove the offset relative to SED-based $L_{\rm bol}$ (see left panel of Fig.~\ref{fig:lbol_comparison}), indicating that dust alone is insufficient and that additional effects are likely required, as BLR changes and/or anisotropic illumination in super-Eddington discs. 

In addition, SEAMBHs, LRDs, and JWST-selected AGN all cluster in the X-ray–weak region of the $L_{\rm bol}/L_{2-10\,\rm keV}$–$L_{\rm bol}$ parameter space (see Fig.~\ref{fig:kbol_lbol}). This distribution is consistent with, although it does not uniquely imply, scenarios in which highly accreting systems produce intrinsically very soft X-ray spectra as a consequence of their high accretion rates. Combined with the fact that, at high redshift, the observed X-ray band probes increasingly higher rest-frame energies, this can lead to a substantial underestimation of the intrinsic X-ray luminosity, naturally accounting for the observed X-ray weakness of JWST-selected AGN and LRDs. Given the current lack of robust constraints on nuclear obscuration for high-redshift sources, deeper and more diagnostic X-ray observations will be required to determine whether the observed X-ray weakness is driven primarily by accretion physics, heavy obscuration, or a combination of both.

Further deep X-ray observations and infrared spectroscopic follow-up are necessary to constrain their physical conditions more robustly. Next-generation X-ray and optical/NIR observatories will be transformative for this field. ESA's  New Advanced Telescope for High-Energy Astrophysics {\it NewAthena} mission, with its Wide Field Imager \citep[WFI;][]{2013arXiv1308.6785R} and large collecting area, will provide an order-of-magnitude gain in sensitivity compared to current facilities and is expected to detect hundreds of AGN at $z>3$, including $\sim$250 at $z>6$ \citep[e.g.][]{Marhesi2020}. Similarly, NASA's proposed {\it Advanced X-Ray Imaging Satellite} \citep[{\it AXIS};][]{2023arXiv231100780R}, offering $\sim$10 times the effective area of {\it Chandra} combined with subarcsecond resolution, would be a game-changer for confirming X-ray suppression in LRDs at the faintest flux levels \citep[e.g.][]{Mushotzky2018,Marhesi2020}. Last but not least, the Extremely Large Telescope (ELT) will be instrumental to find even more LRDs at all Cosmic Times, and in general to better understand the physics of high accretion. 

\begin{acknowledgements}
      AT acknowledges financial support from the Bando Ricerca Fondamentale INAF 2022 Large Grant 'Toward a holistic view of the Titans: multi-band observations of $z>6$ QSOs powered by greedy supermassive black holes' and from the Bando Ricerca Fondamentale INAF 2024 Large Grant 'The DEepest study of LUminous QSOs in X-ray at z=2-7'. AT, GV and MB acknowledge the support of the ESO Scientific Visitor Programme. CR acknowledges support from SNSF Consolidator grant F01$-$13252, Fondecyt Regular grant 1230345, ANID BASAL project FB210003 and the China-Chile joint research fund. GV acknowledges support by European Union’s HE ERC Starting Grant No. 101040227 - WINGS. LCH was supported by the National Science Foundation of China (12233001) and the China Manned Space Program (CMS-CSST-2025-A09). JMW is supported by NSFC-12333003. The authors thank the anonymous referee for the useful suggestions which helped in improving the manuscript.
\end{acknowledgements}

\bibliographystyle{aa} 
\bibliography{biblio} 
\begin{appendix} 

\end{appendix}
\end{document}